\documentclass[fleqn,usenatbib]{mnras}

\usepackage{newtxtext,newtxmath}
\usepackage[T1]{fontenc}

\DeclareRobustCommand{\VAN}[3]{#2}
\let\VANthebibliography\thebibliography
\def\thebibliography{\DeclareRobustCommand{\VAN}[3]{##3}\VANthebibliography}

\usepackage{graphicx}	% Including figure files
\usepackage{amsmath}	% Advanced maths commands

\title[Short title, max. 45 characters]{Machine Learning-Based Classification of Active Galaxies and Estimation of Supermassive Black Hole Masses}
\author[F. Mazoochi]{
F. Mazoochi,$^{1}$\thanks{famazoochi@ipm.ir, spacemazoochie@gmail.com}
R. Karimi,$^{1,2}$
M. H. Zhoolideh Haghighi$^{1,2}$
and F. S. Tabatabaei$^{1,3}$
\\
$^{1}$School of Astronomy, Institute for Research in Fundamental Sciences-IPM, Tehran 19395-5531, Iran\\
$^{2}$Department of Physics, K. N. Toosi University of Technology, Tehran 15875-4416, Iran\\
$^{3}$Max-Planck-Institut f\"ur Radioastronomie, Auf dem H\"ugel 69,Bonn D-53121, Germany
}

\date{Accepted XXX. Received YYY; in original form ZZZ}

\pubyear{\the\year{}}

\begin{document}
\label{firstpage}
\pagerange{\pageref{firstpage}--\pageref{lastpage}}
\maketitle

\begin{abstract}
Distinguishing active galaxies from star-forming galaxies is essential for understanding galaxy evolution. Diagnostic methods like the BPT (Baldwin, Phillips, and Terlevich) diagram use optical emission-line ratios to separate galaxies. However, with growing availability of large surveys and high-resolution instruments, manually identifying galaxy types has become increasingly challenging. In this study, we investigate machine learning to classify active and star-forming galaxies using properties like stellar mass, stellar velocity dispersion, colour, redshift, and [O III] luminosity. These new approaches enable faster AGN/star-forming galaxy classification than the BPT diagram and provide a flexible, scalable alternative that can complement traditional diagnostics, particularly for large surveys or low-quality data. We employ four classification algorithms---Decision Tree, Random Forest, Support Vector Classifier (SVC), and k-Nearest Neighbours (KNN)---using the Galaxy Zoo 1 dataset derived from the SDSS sample. The dataset contains 47,675 galaxies within the redshift range 0.02--0.05, including 17,002 pure star-forming and 2,254 active galaxies, labeled using the BPT diagram. These labels train and evaluate our models through confusion matrices, learning curves, and receiver operating characteristic (ROC) curves. Among the four algorithms, the SVC and Random Forest models achieve the highest accuracy of approximately 93\%, while KNN shows the lowest at 88\%. Furthermore, we estimate supermassive black hole masses using stellar velocity dispersion ($\sigma$) and the $M_{\rm BH}-\sigma$ relation. We apply four regression models---Random Forest Regressor, Support Vector Regressor (SVR), KNN Regressor, and Polynomial Regression. All four models produce similar results, with $R^2$ values from 0.75 to 0.77, indicating consistent performance.
\end{abstract}


%%%%%%%%%%%%%%%%%%%%%%%%%%%%%%%%%%%%%%%%%%%%%%%%%%

%%%%%%%%%%%%%%%%% BODY OF PAPER %%%%%%%%%%%%%%%%%%

\section{Introduction}
Active Galactic Nuclei (AGN) are extremely luminous central regions of active galaxies powered by a non-stellar origin in the core. The presence of accretion disks around supermassive black holes (SMBHs) at the centers of active galaxies is associated with significant radiation at all wavelengths \citep{netzer2013physics}. The immense gravitational and frictional forces heat this disk to millions of degrees, causing it to emit powerful radiation across the entire electromagnetic spectrum, from radio waves to gamma rays. Through this paper, galaxies that host these active nuclei are called AGN. These types of galaxies, based on accretion rate, inclination, and obscuration, can be divided into distinct classes, including quasars, Seyfert galaxies, LINERs, blazars, and radio galaxies \citep{carroll2017introduction}. 

AGNs play a crucial role in physics and the evolution of galaxies.  Through feedback mechanisms, they can regulate star formation by removing or heating cold gas in the interstellar medium of galaxies, which is the main fuel for star formation \citep{mo2010galaxy,li2018stellar}. Moreover, they are believed to have shaped the large-scale structure in galaxy clusters and groups \citep{gitti2012evidence}. Methods of separating pure active and star-forming galaxies have, hence, an important application in the astronomical community. %The separation of these galaxies from star-forming galaxies is crucial for astronomers in their studies. 
Star-forming galaxies are galaxies that are actively producing new stars, typically on timescales of tens to hundreds of millions of years. These galaxies are characterised by the presence of large amounts of cold gas, dust, and young massive stars that emit strongly in the ultraviolet and ionise surrounding gas, producing prominent emission lines in their spectra \citep{mo2010galaxy,carroll2017introduction}. 

%%%%%%%%%%%
To separate these two types of galaxies, astronomers use various methods, such as the BPT (named after Baldwin, Phillips, and Terlevich, \citep{baldwin1981classification}) diagnostic diagram. The BPT diagram has become a standard method in analysing large spectroscopic surveys like the Sloan Digital Sky Survey (SDSS) \citep{york2000sloan}, helping to map the distribution of ionising processes in galaxies and understand galaxy evolution. Some models, such as mid-IR selection techniques \citep{cardamone2008mid} or hard X-ray selection \citep{winter2008x}, are inefficient in characterising AGNs as they miss a large number of genuine AGNs or are confirmed to a small area in the SDSS. AGNs can produce radio emission unrelated to star formation, leading to lower $q_{\mathrm{IR}}$ values compared to star-forming galaxies and enabling their separation. However, this method misses radio-quiet AGNs and is sensitive to redshift and K-correction uncertainties.

The BPT diagram distinguishes star-forming galaxies from AGN based on the four optical line ratios [O III]/H$\beta$ versus [N II]/H$\alpha$, [S II]/H$\alpha$, and [O I]/H$\alpha$ \citep{kewley2006host}. These line ratios are sensitive to the hardness of the ionising radiation field. Star formation produces a softer radiation field (from young, massive stars), while AGN produce harder, high-energy radiation (from accreting black holes), resulting in different positions on the diagram. Star-forming galaxies occupy the lower-left part of the diagram, and AGN (Seyfert galaxies and LINERs) fall in the upper-right region. LINER (Low-Ionisation Nuclear Emission-line Regions) are a class of galaxies or galactic nuclei characterised by strong optical emission lines from weakly ionised or neutral atoms, such as [OI], [NII], and [SII] \citep{ho1999liners}. Seyfert galaxies show strong emission lines in their optical spectra, originating from gas ionised by the central black hole’s radiation \citep{Osterbrock1989}. \citet{kewley2001theoretical} and \citet{kauffmann2003host} introduced a theoretical and empirical separation line that are called Ke01 and Ka03. Ke01 is the maximum starburst line based on theoretical models, and Ka03 is the empirical line that divides pure star-forming galaxies from AGN. The galaxies lying between these two lines are composite galaxies that are both AGN and star-forming galaxies \citep{kewley2006host}.

Nevertheless, separating AGNs from star-forming galaxies using optical line ratios and large spectroscopic surveys to construct the BPT diagram is not always possible. One possible solution for this problem could be using machine learning techniques, which have been shown to be successful in different
branches of astronomy/astrophysics. 
%In addition, the new machine learning approaches have been shown to be successful in different branches of astronomy/astrophysics.
For instance, people have applied diverse machine learning algorithms to classify various types of stars \citep{ghaziasgar2025dusty,li2025machine} or the morphology of galaxies in large datasets \citep{de2004machine,Chen_2025}. The detection of exoplanets has been complicated recently by machine learning \citep{malik2022exoplanet,karimi2025machine}. When the photometric and spectroscopic measurements are unavailable, the existing machine learning models are used to estimate redshift \citep{janiurek2024transferability}. Moreover, gravitational microlensing events in gamma-ray bursts have been identified using machine learning \citep{haghighi2025machine}. Furthermore, deep learning has been shown to be very effective for protecting ground-based observations \citep{haghighi2025deep}. Therefore, with the aid of sophisticated machine learning approaches, we can identify galaxies more quickly than in the past. The supervised machine learning models can classify galaxies based on their features and the labels provided by previous surveys \citep{sadeghi2021morphological,zeraatgari2024machine}. Numerous studies have recently been interested in applying traditional and physical classifications of AGNs and star-forming galaxies, such as the BPT diagram and so on, in faster techniques like machine learning in large surveys. For example, \cite{teimoorinia2018discrimination} applied a machine learning approach that uses the fluxes and equivalent widths of [O III] and H$\beta$, the $\rm D_n$(4000) index, galaxy colours, and stellar mass to distinguish between the two classes when H$\alpha$ and [N II] measurements are not available. Similarly, a supervised classifier was applied to deep-field radio continuum data from the Low-Frequency Array (LOFAR) survey, using multi-wavelength photometry and SED-based labels to identify AGN and star-forming galaxies \citep{karsten2023multi}. Furthermore, \cite{silima2025machine} studied radio sources from the MeerKAT International GHz Tiered Extragalactic Exploration Survey (MIGHTEE) and used features such as the infrared–radio correlation parameter, optical compactness, MIR colours, and others. This study shows that classification performance improves through the combination of features \citep{silima2025machine}.

We use the Galaxy Zoo data \citep{schawinski2010galaxy} to apply classification models for two classes: Star-forming and AGN galaxies. We classify galaxies that are purely AGN or purely star-forming. In addition to distinguishing AGNs from star-forming galaxies, we apply regression algorithms to predict one of the AGN properties, the mass of the SMBH ($\rm M_{BH}$). This property also plays a significant role in the study of the growth and evolution of galaxies. The mass of the SMBH can be indirectly estimated from the stellar velocity dispersion \citep{gitti2012evidence}, a relation that we will further discuss in this paper.

%organization of the paper
This paper is organised as follows:
In Section~\ref{sec:data}, we describe the dataset used in this study. The applied models to classify galaxies are introduced in Section~\ref{sec:model}. In Section~\ref{sec:result}, the evaluation methods and classification results are presented with comparisons of the performance of different models. The regression models and their results for predicting the masses of SMBHs in AGNs are presented in Section~\ref{sec:regression}. Finally, the summaries and conclusions of both the classification and regression analyses are provided in Section~\ref{summary}.

%See \texttt{mnras\_sample.tex} for a more complex example, and \texttt{mnras\_guide.tex}for a full user guide\citet{Fournier1901},and describes the problem the authors aim to solve \citep[e.g.][]{vanDijk1902} \citet{deLaguarde1903, delaGuarde1904}.
%Refer back to them as e.g. equation~(\ref{eq:quadratic}).

%Figures are referred to as e.g. Fig.~\ref{fig:example_figure}, and tables as e.g. Table~\ref{tab:example_table}.

\section{Data}\label{sec:data} 
The data used in this study are presented in the Galaxy Zoo 1 (GZ1) paper on AGN host galaxies \citep{schawinski2010galaxy}. The GZ1 (launched July 2007) is a project where members of the public help classify galaxies based on their shapes in astronomical images \citep{lintott2008galaxy}. The morphology of nearly a million galaxy images from the Sloan Digital Sky Survey (SDSS, \citet{york2000sloan}) is classified by the GZ1 project. This dataset comes from the clean sample, where 80\% of the majority agree on the GZ1 morphology of any object \citep{land2008galaxy} with the emission line classifications, stellar masses, and velocity dispersions \citep{lintott2008galaxy,lintott2011galaxy}. In this dataset, the galaxies in redshift of 0.02 to 0.05 ($0.02<Z<0.05$) and magnitude of z band less than -19.5 ( $\rm M_z < -19.5$ AB mag) are selected. The z band is chosen rather than the r band since the redder z band emanates from the older, more massive stellar population, making it a more accurate indicator of stellar mass \citep{schawinski2010galaxy}. This yields a total of 47,675 galaxies. The photometric and spectroscopic data used in this survey are taken from the SDSS-DR7 \citep{york2000sloan,strauss2002spectroscopic,abazajian2009seventh}.

\subsection{Feature Selection} \label{sec:feature}

Here is a brief explanation of the features used in this study from the dataset:

\begin{itemize}
    \item {\bf Redshift:} The redshift ($z$) in this dataset ranges from 0.02 to 0.05 and is obtained from SDSS spectra classified as {\it GALAXY} \citep{strauss2002spectroscopic}.
 
    \item {\bf Velocity dispersion:} The stellar velocity dispersion ($\sigma_{\star}$) quantifies the spread in stellar velocities within a galaxy or stellar cluster and is used to study galaxy dynamics and masses. This feature is measured for the stellar population near the black hole. The mass of an SMBH is obtained using the well-known $M_{\rm BH}-\sigma_{\star}$ relation \citep{gebhardt2000relationship}. We discuss more about this relation in Section\,\ref{sec:regression}.

    \item {\bf Stellar Mass:} The stellar masses ($\rm M_{\star}$) in this sample are estimated by fitting the five SDSS photometric bands to star formation history models \citep{schawinski2010galaxy}. This parameter is fundamental for understanding galaxy formation and evolution. In this dataset, the logarithm of the stellar mass, expressed in solar mass ($\rm M_{\odot}$), is used.

    \item {\bf Luminosity of [OIII]:} This forbidden line emission ($\rm L_{[OIII]}$) is a key parameter for studying the accretion state of AGNs and the growth rate of their central black holes. The $\rm [OIII]\ \lambda 5007$ flux is measured for each processed SDSS spectrum, and the total luminosity, $\rm L_{[OIII]}$, is calculated after correcting for extinction based on the measured Balmer decrement \citep{schawinski2010galaxy}.
    
    \item {\bf Color:} In this study, this property is defined using the SDSS filters and corresponds to the difference between magnitudes in two filters. Specifically, we consider the colors $u-g$, $g-r$, and $r-i$.

\end{itemize}

\subsection{AGN Selection \&  Classification Target Labels} 

The main important part of this dataset is AGN selection, as it is a challenging problem. The emission-line selection is used for this dataset to characterize galaxies. For this aim, emission-line fluxes of [OIII]/H$\beta$, OI/H$\alpha$, and [NII]/H$\alpha$ are measured using an analysis tool called Gas AND Absorption Line Fitting algorithm (GADALF, \citet{sarzi2006sauron}). The pure star-forming galaxies are distinguished by the theoretical Ke01 line, and objects between this extreme starburst line and the empirical Ka03 are labeled as the composite galaxies where both AGNs and star formation are comparable in ionizing luminosity. The left galaxies, where emission lines are dominated by sources of ionization other than young stars, are AGNs. These non-stellar sources are empirically derived from two sources, obviously in the [OI]/H$\alpha$ diagram. The lower branch of this propagation is LINERs, and the upper branch is Seyfert (Type 2 with narrow line emission) AGNs.

The number and percentage of galaxies based on the emission lines and BPT classification in our dataset are reported in Table~\ref{tab:number}. In this dataset, we discarded no emission and composite galaxies. By excluding the composite class in this study, we acknowledge that we might be omitting an important stage in AGN–galaxy co-evolution at low redshift. The main challenge with this class lies in the absence of readily available spectral diagnostics that would allow a clean separation of AGNs and star formation activity.

\begin{table*}
    \centering
        \caption{Numbers and percentages of BPT class labels in our dataset.}
    \begin{tabular}{c c c c c c}
    \hline \hline
       BPT class & No emission  & Star-forming & Composite & Seyfert & LINER\\
       \hline
       Number &23729&17002&4690&942&1312\\ 
       Percentage (\%) &49.77&35.66&9.84&1.97&2.75 \\
    \hline 
    \end{tabular}
    \label{tab:number}
\end{table*}

We noted that 6737 data points reported identical velocity dispersion value, corresponding to the minimum measurable dispersion, which lies below the detection limit of the SDSS spectrograph. To enhance the training results, we excluded these points from the dataset. To classify pure star-forming and AGN galaxies, we combine the remaining two Seyfert and LINER galaxies. This combination results in 2181 AGN samples, which is fewer than the remaining 10338 star-forming samples. To prevent skewed data issues and keep a balance between these two labels \citep{chawla2004special}, we randomly selected 2181 star-forming galaxies and then applied machine learning methods.

%-------------------------------------------------

\section{Machine Learning Classification Models}\label{sec:model}
Various models are defined to train features and targets and predict the new values after training the models. In this study, we used a selection of these models to classify star-forming and AGN galaxies, as described in the following subsection. 

\subsection{Model definition \& training} 
The models we used in this study are:
\begin{itemize}

   % \item {\bf Logistic Regression classifier:} This method predicts the probability that something belongs to a particular category based on using \textbf{Sigmoid function} ($\sigma(z)=1/(1+e^{-z})$) \citep{wright1995logistic}. The sigmoid function maps any real-valued number into a value between 0 and 1, which can be interpreted as a probability.

    \item {\bf Decision Tree Classifier:} This supervised Learning algorithm uses a sequence of {\it if-then-else} to split data into branches based on feature values. The final split data corresponds to predicted classes \citep{safavian1991survey}. The decision tree model is easy to interpret and fast to train and predict, but prone to overfitting.   
    
    \item {\bf Random Forest Classifier:} An ensemble machine learning model that makes predictions by combining the outputs of many decision trees, each trained on slightly different versions of the data \citep{breiman2001random}. Although this classifier model is slower than a single decision tree in training and predicting, it reduces overfitting and is more accurate and robust.   

    \item {\bf Support Vector Machine Classifier (SVC):} This method classified data by the optimal hyperplane to different classes \citep{cortes1995support}. The SVC maximizes the margin, which is the distance between the hyperplane and the nearest data points (the support vectors) from each class. A large margin generally leads to a more robust classifier. This algorithm is robust against overfitting and effective in high-dimensional data with a clear margin. On the other hand, it struggles with noise and overfitting classes, and training is slow for large datasets.  

    \item {\bf k-Nearest Neighbors (KNN) Classifier:} It's a simple, non-parametric method that classifies a new data point based on the majority class of its nearest neighbors in the feature space. The new data point is assigned the class that is most common among its "K" nearest neighbors \citep{cunningham2021k}.
    This model is easy to implement, but computationally expensive for large data and sensitive to irrelevant features and the "K" parameter.

\end{itemize}
It is worth mentioning that we use 80\% of the data for training and 20\% for testing.

%--------------------------------------------------

\section{Classification Results}\label{sec:result} %
In this section, we presented and discussed various metrics and approaches used to evaluate the applied classifier and compare their results. Next, we highlight the methods with the strongest and weakest performance in classifying our galaxy sample.

%\subsection{Evaluation Metrics}
To evaluate the performance of our models, we first use several metrics, including accuracy, precision, recall, and the F1-score \citep{baldwin1981classification,van1986non}. %Accuracy measures the proportion of correct predictions out of all predictions. 
The accuracy is the ratio of accurate predictions to total predictions. As this metric can't be precise enough to report the validation of models, the precision, recall and F1-score \citep{christen2023review} are used in addition to the accuracy. %Precision and recall are the ratios of true positives to total predicted positives and total actual positives. The F1-score is the harmonic mean of these two metrics (F1-score = 2 Precision $\times$ Recall / (Precision+Recall)). The F1-score is particularly useful when dealing with imbalanced datasets. 
The bar plot below presents the results of these metrics for our trained models. 

\begin{figure}
    \centering
    \includegraphics[width=\linewidth]{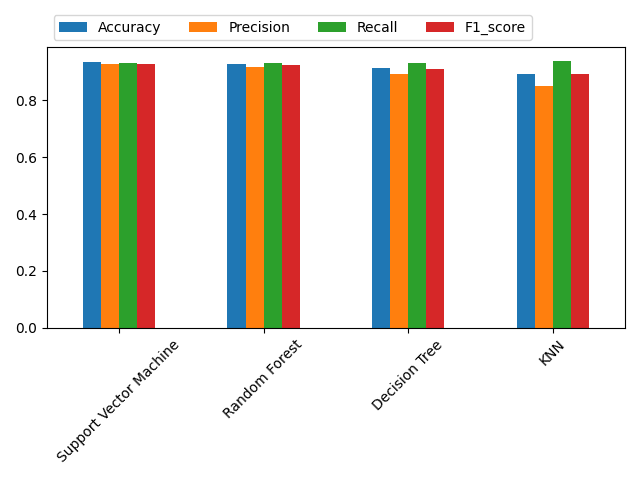}
    \caption{Bar plots of evaluation metrics used for classification models.}
    \label{pic:validation}
\end{figure}
Table~\ref{tab: evaluation} presents the quantitative evaluation of the four classification models in terms of accuracy, precision, recall, and F1-score. 

\begin{table}
    \centering
    \caption{Values of evaluation metrics for classification models.}
\begin{tabular}{c c c c c}
        \hline\hline
        Model & Accuracy & Precision & Recall & F1-score \\
        \hline
        \multicolumn{5}{c}{Down-Sampling}\\
        \hline 
         SVC & 0.932 & 0.927 & 0.929 & 0.928 \\
         Random Forest & 0.928 & 0.918 & 0.929 & 0.924 \\
         Decision Tree & 0.913 & 0.890 & 0.929 & 0.909 \\
         KNN & 0.892 & 0.848 & 0.939 & 0.891 \\
         \hline
         \multicolumn{5}{c}{SMOTE}\\
         \hline
         SVC & 0.932 & 0.765 & 0.914 & 0.833 \\
         Random Forest & 0.917 & 0.716 & 0.916 & 0.804 \\
         Decision Tree & 0.910 & 0.703 & 0.895 & 0.788 \\
         KNN & 0.901 & 0.680 & 0.886 & 0.770 \\
         \hline
         \multicolumn{5}{c}{Stratified Loss}\\
         \hline
         SVC & 0.935 & 0.758 & 0.912 & 0.828 \\
         Random Forest & 0.938 & 0.770 & 0.912 & 0.835 \\
         Decision Tree & 0.923 & 0.726 & 0.887 & 0.798 \\
         KNN & 0.931 & 0.820 & 0.767 & 0.792 \\
        \hline
        \end{tabular}
    \label{tab: evaluation}
\end{table}

Among them, the SVC demonstrates the strongest overall performance, achieving the highest values across all metrics with an accuracy of 0.932, precision of 0.927, recall of 0.929, and F1-score of 0.928. These results indicate that this model is highly reliable in classifying star-forming galaxies and AGN. The Random Forest also performs strongly, with an accuracy of 0.928, precision of 0.918, recall of 0.929, and F1-score of 0.924, showing only slightly lower performance than the SVC. The Decision Tree achieves an accuracy of 0.913, precision of 0.890, recall of 0.929, and F1-score of 0.909. While its recall suggests that it can recover most true labels, the lower precision and F1-score reflect a higher rate of misclassification relative to the SVC and Random Forest. Finally, the KNN model records the lowest overall performance, with an accuracy of 0.892, precision of 0.848, recall of 0.939, and F1-score of 0.891. The high recall shows that KNN is effective at identifying true positives, particularly AGN, but its comparatively low precision highlights a greater tendency toward false positives.

To address class imbalance and prevent model bias, we compared random selection and down-sampling against some robust techniques, including SMOTE (Synthetic Minority Over-sampling Technique), and stratified loss functions. SMOTE balances a dataset by generating new synthetic examples for the minority class, whereas stratified loss balances the model's focus by mathematically penalizing errors on those rare examples more heavily during training. Our results indicate that the overall classification accuracy remains comparable across all three methods. However, we find that the F1-scores obtained using SMOTE and stratified loss are slightly lower than those achieved with down-sampling. This suggests that, for our specific dataset, down-sampling does not significantly degrade performance despite reducing the number of training samples.

\begin{figure}
    \centering
        \includegraphics[width=\columnwidth]{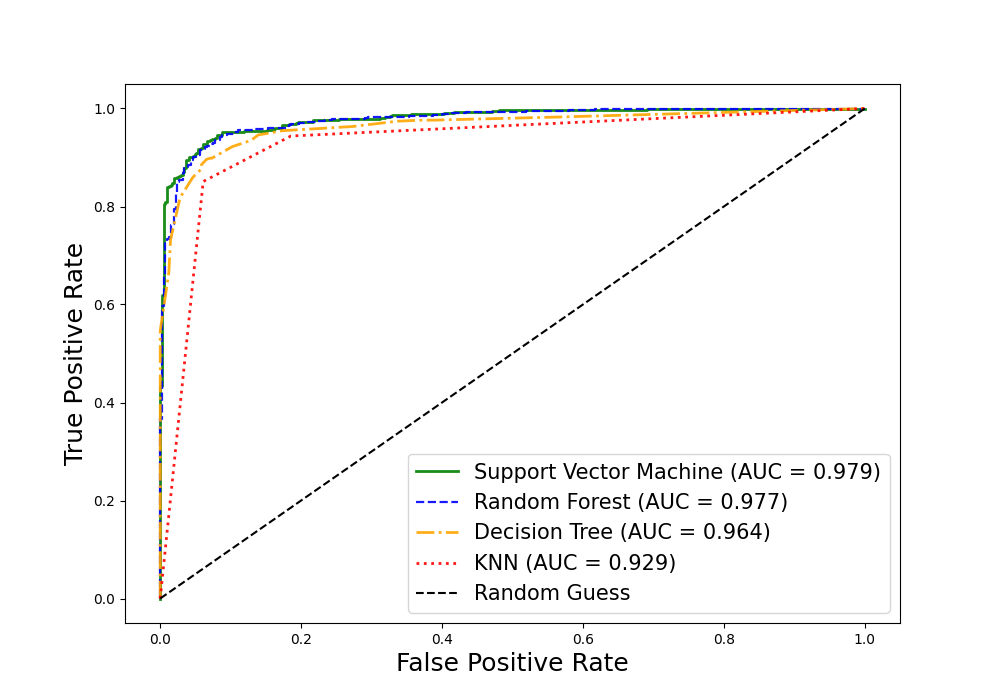}

    \caption{ROC curves for applied classification models. The SVC model indicates the best performance (AUC=0.98), and the KNN model shows the worst one (AUC=0.93).}
    \label{fig:roc_curve}
\end{figure}

%\subsection{ROC Curve}
In addition to the mentioned evaluation metrics, we plotted ROC curves. The ROC curve (Receiver Operating Characteristic curve) is used to evaluate the performance of a binary classifier by showing the trade-off between true positive rate (TPR) and false positive rate (FPR) at different classification thresholds \citep{baron2019machine}. Fig.~\ref{fig:roc_curve} indicates ROC curves for our classified models and the corresponding Area Under the Curve (AUC) values, which further validate these findings. The SVC demonstrates the highest discriminative ability with an AUC of 0.979, followed closely by Random Forest (0.977)  and Decision Tree (0.964). The KNN model yields the lowest AUC (0.929), consistent with its weaker overall performance.

\begin{figure*}
    \centering
       
        \includegraphics[width=0.48\textwidth]{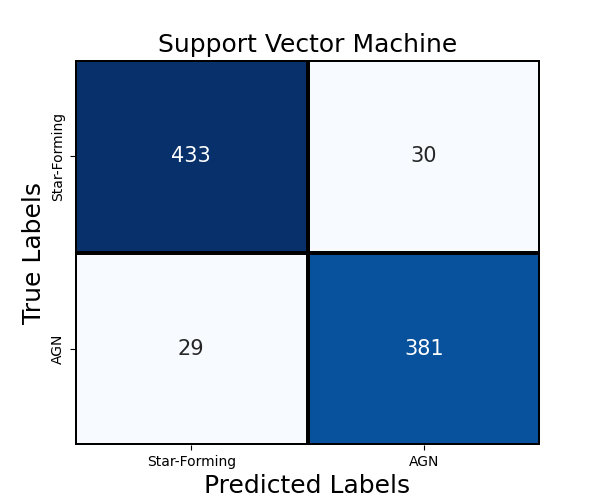}
        \includegraphics[width=0.48\textwidth]{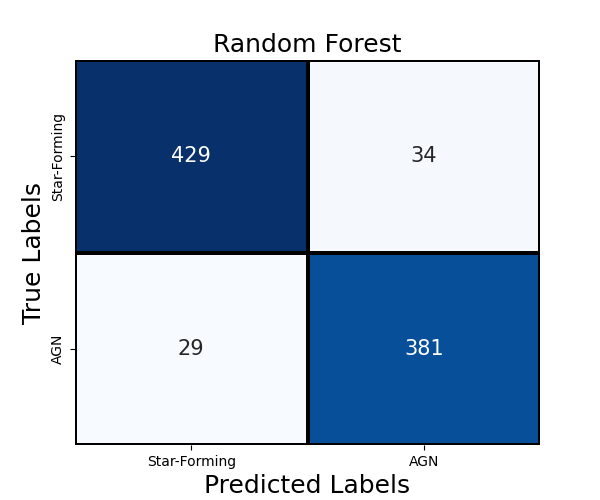}

        \includegraphics[width=0.48\textwidth]{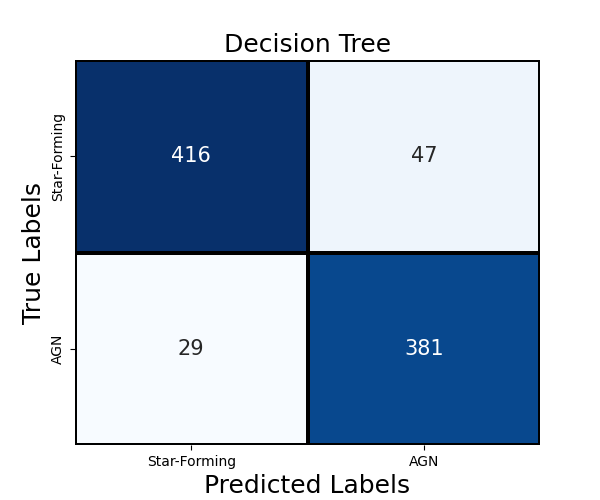}
        \includegraphics[width=0.48\textwidth]{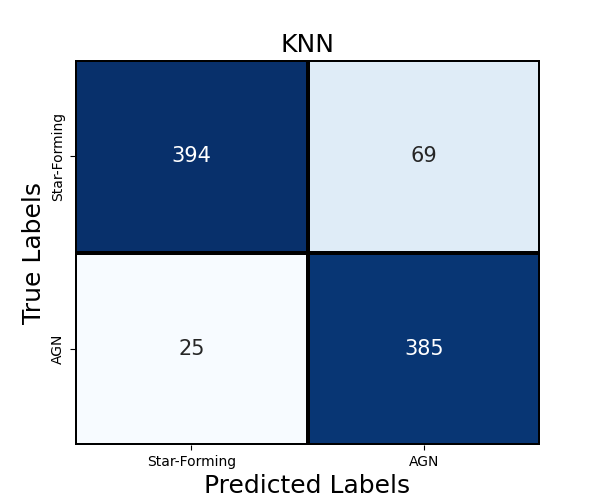}
    \caption{Confusion matrix results for our classification models. The SVC and Random Forest classifiers achieve the highest number of correctly predicted class labels. The misclassification rates increase in the Decision Tree model. The KNN classifier produces the most false predictions in the Star-Forming class, indicating the weakest performance among the four classifiers.}
    \label{fig:confusion_matrix}
\end{figure*}

%\subsection{Confusion Matrix}
In this work, we also show the confusion matrix, which is a table that summarizes how well a classification model performs by showing the counts of correct and incorrect predictions. The ideal classification yields a diagonal matrix, where only the diagonal elements are non-zero \citep{baron2019machine}.
The confusion matrix provides additional insight by detailing predictions in terms of true positives, true negatives, false positives, and false negatives. In Fig.~\ref{fig:confusion_matrix}, confusion matrices illustrate how effectively each model distinguishes between star-forming galaxies and AGN. The SVC correctly classifies 433 star-forming galaxies and 381 AGN, with only 29 false star-forming and 30 false AGN, underscoring its robustness in separating the two populations with high accuracy. The Random Forest also performs strongly, correctly identifying 429 star-forming galaxies and 381 AGN, while reducing misclassifications to 29 and 34, respectively. The Decision Tree delivers solid but less accurate results, with 416 star-forming galaxies and 381 AGN being correctly classified, alongside 29 false star-forming and 47 false AGN, indicating higher misclassification rates. The KNN classifier performs comparatively less well, correctly identifying 394 star-forming galaxies and 385 AGN, but producing 25 false star-forming and 69 false AGN. While it achieves strong performance in detecting AGN, it is more prone to errors in identifying star-forming galaxies relative to the other models.

\begin{figure*}
    \centering
       
        \includegraphics[width=0.48\textwidth]{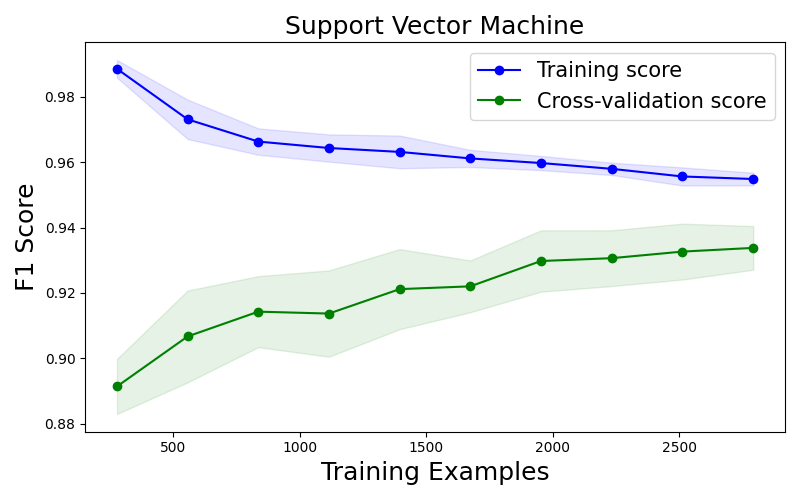}
         \includegraphics[width=0.48\textwidth]{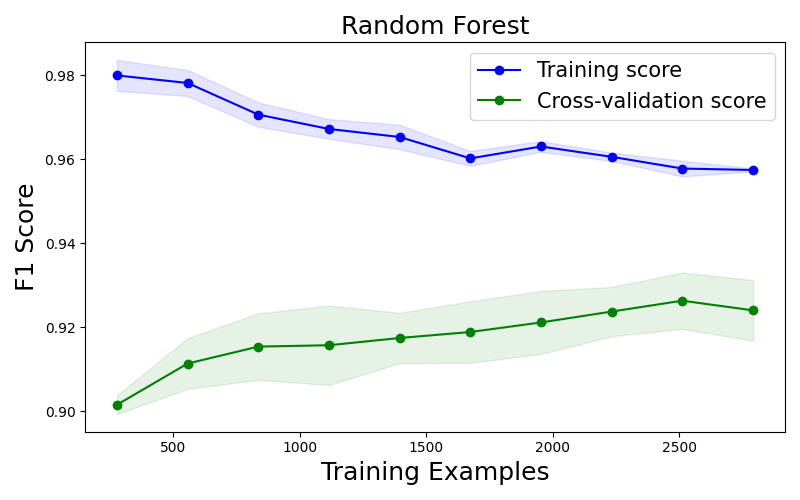}

        \includegraphics[width=0.48\textwidth]{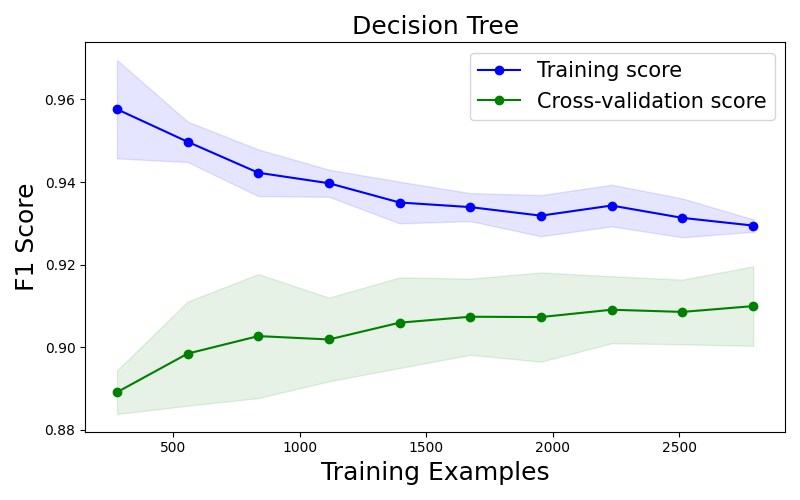}
        \includegraphics[width=0.48\textwidth]{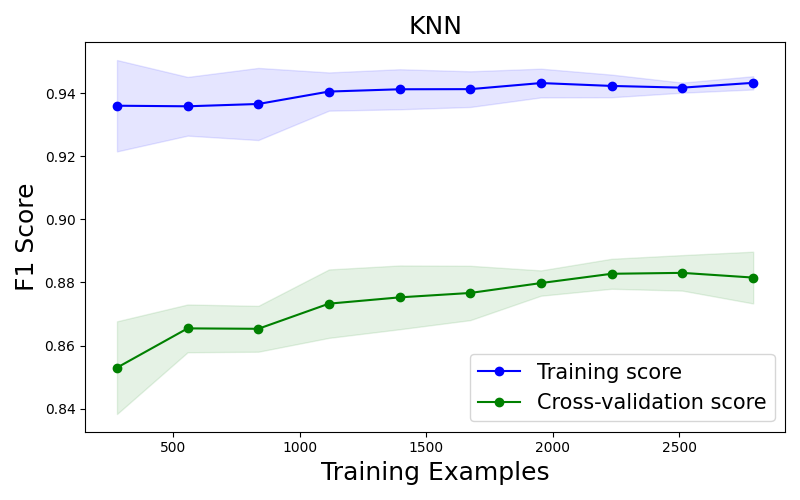}
    \caption{Learning curve results (F1-score versus training examples) for the four classifier models. The blue line denotes the training score, while the green line represents the cross-validation score. The shaded regions around each curve indicate the standard deviation.} 
    \label{fig:learning_curve}
\end{figure*}

%\subsection{Learning Curve}
Another method used to assess the performance of the models is the learning curve. This curve is a plot that shows how a model’s performance changes over time (or with more training data). It’s often used to understand whether your model is learning effectively or if it’s underfitting/overfitting \citep{perlich2011learning}. Fig.~\ref{fig:learning_curve} presents the learning curves for the four models, showing the evolution of training and cross-validation F1-scores as the number of training examples increases. These curves provide key insights into model bias, variance, and generalization performance. The SVC demonstrates strong behavior, with training scores starting near 0.99 and stabilizing around 0.96, while validation scores steadily rise to $\sim$0.93. The narrow gap between the curves indicates excellent generalization and minimal overfitting, confirming that SVC achieves a reliable balance between model complexity and predictive accuracy. The Random Forest also shows strong generalization, with training scores decreasing gradually from near 0.98 to 0.96 and validation scores improving to around 0.92. The small gap between curves highlights its robustness and effective balance between bias and variance. The Decision Tree achieves high training scores ($\sim$0.94) but lower validation performance (0.91). %, reflecting overfitting. %While validation improves slightly with more data, the persistent gap underscores its limited generalization ability.
For all three cases, the trends show that by adding more data, the training and validation curves will get closer together and lead to better results. In contrast, the KNN classifier displays the weakest performance, with training scores plateauing near 0.94 and validation scores remaining lower ($\sim$0.87). The wide gap between curves indicates persistent overfitting, where the model memorizes training examples but struggles to generalize to unseen data. Although validation performance improves slightly with larger datasets, KNN’s limitations in handling complex decision boundaries make it less reliable than the other models.

Overall, the collective results from performance metrics, ROC–AUC values, confusion matrices, and learning curves consistently demonstrate that the SVC and  Random Forest classifiers are the most effective and reliable models for distinguishing between star-forming galaxies and AGN.

%\subsection{Feature Importance} \label{appendix:1}

\begin{figure*}
    \centering
    \includegraphics[width=0.7\textwidth]{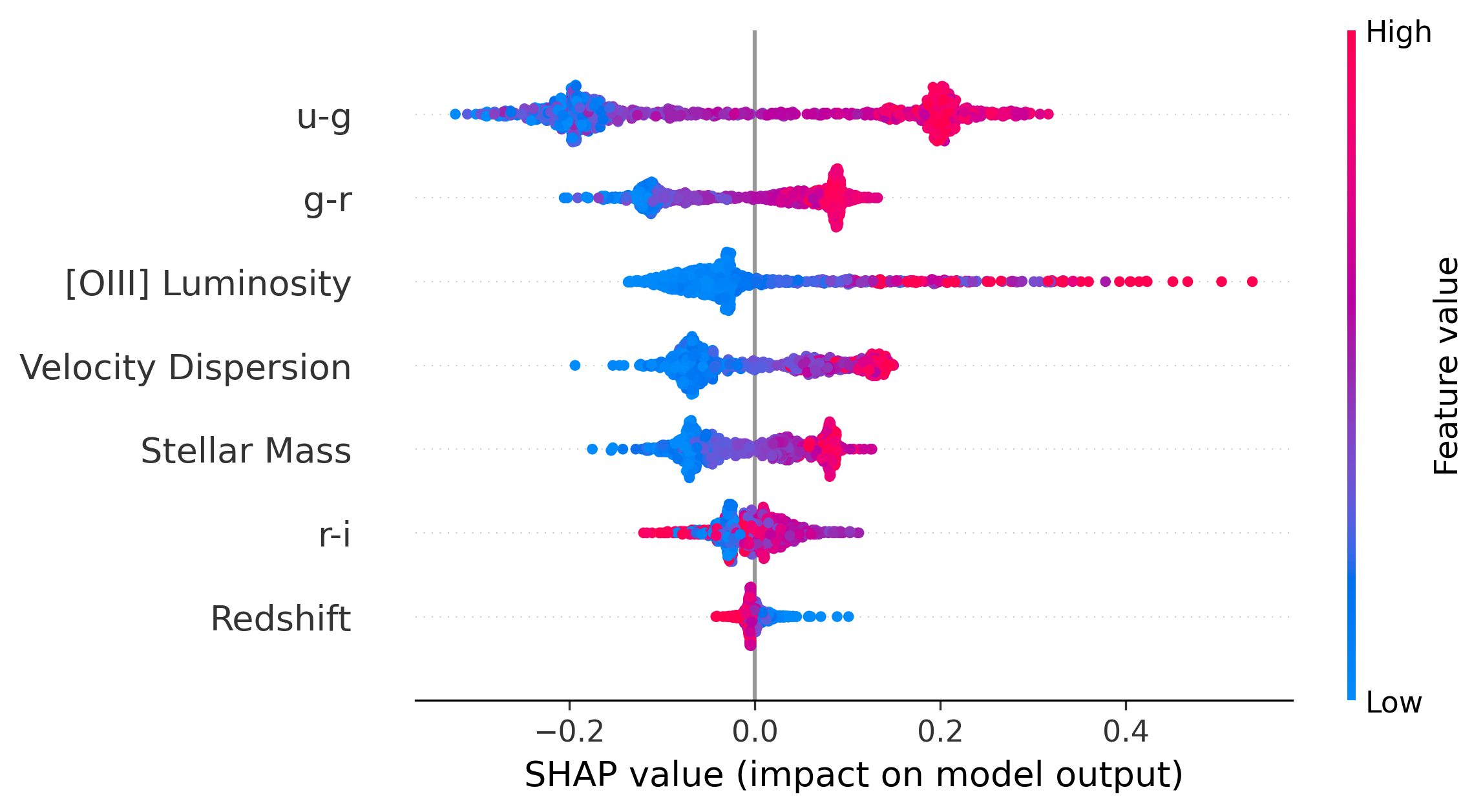}
    \caption{SHAP values are shown for each feature, with the color bar representing the actual values of the corresponding features.}
    \label{pic:shap}
\end{figure*}

\begin{figure*}
    \centering
    \includegraphics[width=0.7\textwidth]{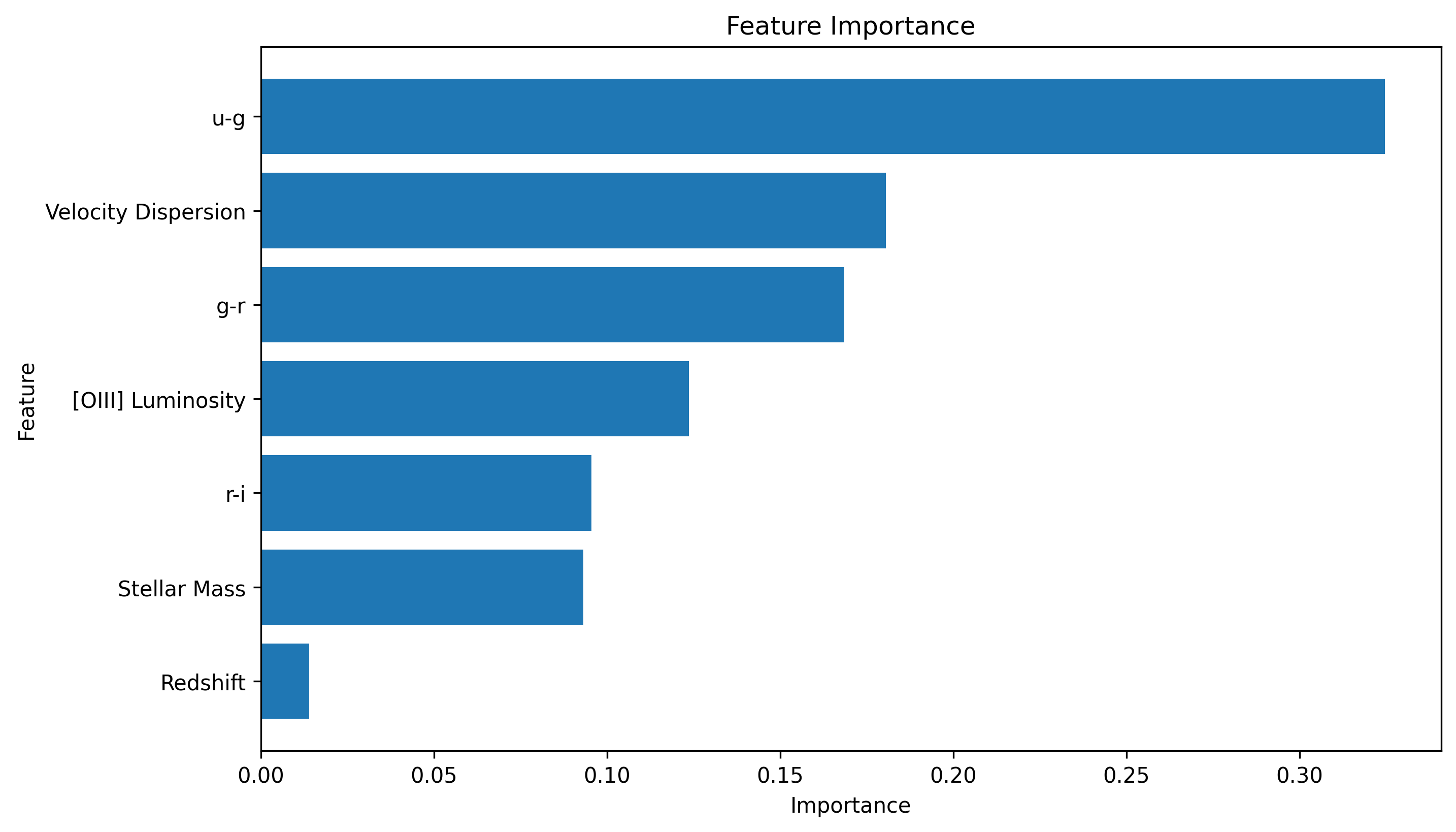}
    \caption{Barplot of feature importance of Random Forest model for individual features.}
    \label{pic:importance}
\end{figure*}

Furthermore, we assess the contribution of each individual feature to the predictions made by the Random Forest classifier. %, which is among the best-performing models in this study. 
To this aim, we present the SHAP (SHapley Additive exPlanations) plot in Fig.~\ref{pic:shap}. This approach provides an interpretation of the model’s predictions by quantifying the impact of each feature. The SHAP value indicates how much a given feature influences a particular prediction—positive values drive the prediction toward higher probabilities of AGN class, while negative values lower it \citep{lundberg2017unified}. The color bar in Fig.~\ref{pic:shap} uses color (red for high, blue for low) to represent the original feature values for each sample. From this visualization, we observe that, apart from redshift, higher feature values generally shift the prediction toward the AGN class. In addition, we display the feature importance as a barplot in Fig.~\ref{pic:importance}. The bar plot reveals that the u–g color and redshift have the highest and lowest importances, respectively.            

\section{Prediction of Black Hole Mass of AGNs} \label{sec:regression}

Most AGNs are too distant for direct measurement of some of their host galaxy properties such as the mass of SMBHs ($\rm M_{BH}$). For this reason, after classifying AGN and star-forming galaxies,  we are interested in predicting $\rm M_{BH}$ at the nuclei of active galaxies. There are various methods to estimate this property. One of the well-known methods is an empirical power-law relation between the mass of SMBH and stellar velocity dispersion, called $\rm M_{BH}-\sigma$ relation \citep{ferrarese2000fundamental,gebhardt2000relationship}.  It suggests a strong link between the growth of SMBHs and the evolution of their host galaxies. This relation is often used to estimate black hole masses in galaxies where direct dynamical measurements are not possible. We should note that the $\rm M_{BH}-\sigma$ relation is robustly defined for elliptical and the bulge of galaxies. For this reason, this relation can exhibit more scatter in star-forming galaxies, which are typically disk-dominated, and is more accurate for quiet (non-star-forming) galaxies. Therefore, we confined our dataset to pure 2254 AGNs. As we mentioned earlier, some data points with the same stellar velocity are discarded before training, and the size of the AGN dataset is reduced to 2181. We use the following formula to convert $\sigma$ (see Section~\ref{sec:data}) to $\rm M_{BH}$ in units of $\rm M_{\odot}$\citep{graham2011expanded}:

\begin{equation}\label{eq:m-sigma}
\rm \rm Log(M_{BH}/M_{\odot}) = 8.15+5.95 \times Log[\sigma/200\, km\,s^{-1}]
%\rm M_{BH} = \alpha (\sigma/\sigma_0)^{\beta},   
\end{equation}

%where $\sigma_0$ is 200\,Km\,s$^{-1}$ and $\rm \alpha = (1.66\pm0.24)\times 10^8\,M_{\odot}$, $\beta = 4.86\pm0.43$ \cite{carroll2017introduction}. 

Therefore, we adopted $\rm M_{BH}$ as the target for regression, while the other introduced features (see Section\,\ref{sec:feature}), together with the BPT class labels, were used as input features for training the regression models. We applied four regression algorithms to predict the logarithm of the $\rm M_{BH}$ ($\rm Log(M_{BH}/M_{\odot})$) for AGNs, derived from equation~(\ref{eq:m-sigma}). It is worth mentioning that the goal of this work is not to derive a new black hole mass relation, but rather to demonstrate that machine learning methods can successfully recover the standard estimates. The regression target values are scaled to the maximum value. Four regression models that we use are as follows :

\begin{itemize}
    \item {\bf Random Forest Regressor:} This algorithm, like the Random Forest Classifier introduced in Section~\ref{sec:model}, builds multiple decision trees during training and merges their predictions to get a more accurate and stable result \citep{loh2011classification}. Although training many trees can be slow, the averaging of multiple trees reduces the risk of overfitting and improves the accuracy.
    
    \item {\bf Support Vector Regressor (SVR):} A Support Vector Regressor (SVR) is an adaptation of the SVC algorithm, which is primarily used for classification (see Section~\ref{sec:model}). While SVR is robust to outliers, it is computationally intensive and requires careful hyperparameter tuning, especially with high-dimensional data.
    
    \item {\bf KNN Regressor:} This method emanates from the KNN classifier, but instead of voting for the nearest class labels, it takes the average (or weighted average) of the "K" neighbors’ values. The KNN regressor performs very well for small samples, but it is sensitive to outliers.
    
    \item {\bf Polynomial Regression:} Eventually, we apply a Polynomial Regression for regression and prediction of $\rm M_{BH}$. This model combines linear regression with additional polynomial features to fit a curve to the data instead of a straight line. This approach allows it to capture non-linear relationships between variables. It's important to choose the polynomial degree carefully, as high-order polynomials can result in overfitting. In this paper, we used the degree of 2.
\end{itemize}

As in the classification task, we used 80\% of the data as the training set \((X_{\text{train}}, y_{\text{train}})\) and the remaining 20\% as the test set \((X_{\text{test}}, y_{\text{test}})\) to estimate $\rm M_{BH}$ in AGNs.

To evaluate the regressor models, we calculated Root Mean Squared Error (RMSE) and R-squared ($\rm R^2$) metrics \citep{chicco2021coefficient} and reported them in Table\,\ref{tab:regression}. $\rm R^2$ represents the proportion of the total variance in the target variable that is explained by the model’s predictions. The value of this metric is between 0 and 1. A higher $\rm R^2$ value suggests a better fit of the model to the observed data. 

To verify that excluding data points with identical stellar velocity values does not affect our findings, we repeat the analysis using the full dataset whose results are reported as RMSE$_{\star}$ and $\rm R^2_{\star}$ in Table\,\ref{tab:regression}. The results change only marginally, since only about $\sim3\%$ of AGNs share the same $\sigma$.

\begin{table}
    \centering
    \caption{Values of the two evaluation metrics ($\rm R^2$ and RMSE) for the regression models applied to estimate $\rm M_{BH}$ in active galaxies. The $\rm R^2_{\star}$ and RMSE$_{\star}$ values are computed using all data points, without excluding those that share the same $\sigma$.}
        \begin{tabular}{l|rrrr}
        \hline \hline
        %&&\multicolumn{2}{*}{Discard same $\sigma$}&\multicolumn{2}{*}{All the data}\\
        Model & $ R^2 $& RMSE&$ R_{\star}^2 $& RMSE$_{\star}$ \\
        \hline
        Random Forest Regressor & 0.774& 0.062&0.792&0.064\\
        SVR &  0.764& 0.064&0.785&0.065\\
        KNN Regressor &  0.752 & 0.065&0.758&0.069\\
        Polynomial Regression &  0.749& 0.066&0.766&0.068\\
        \hline
        \end{tabular}
    
    \label{tab:regression}
\end{table}

\begin{figure*}
    \centering
    \includegraphics[width=0.8\textwidth]{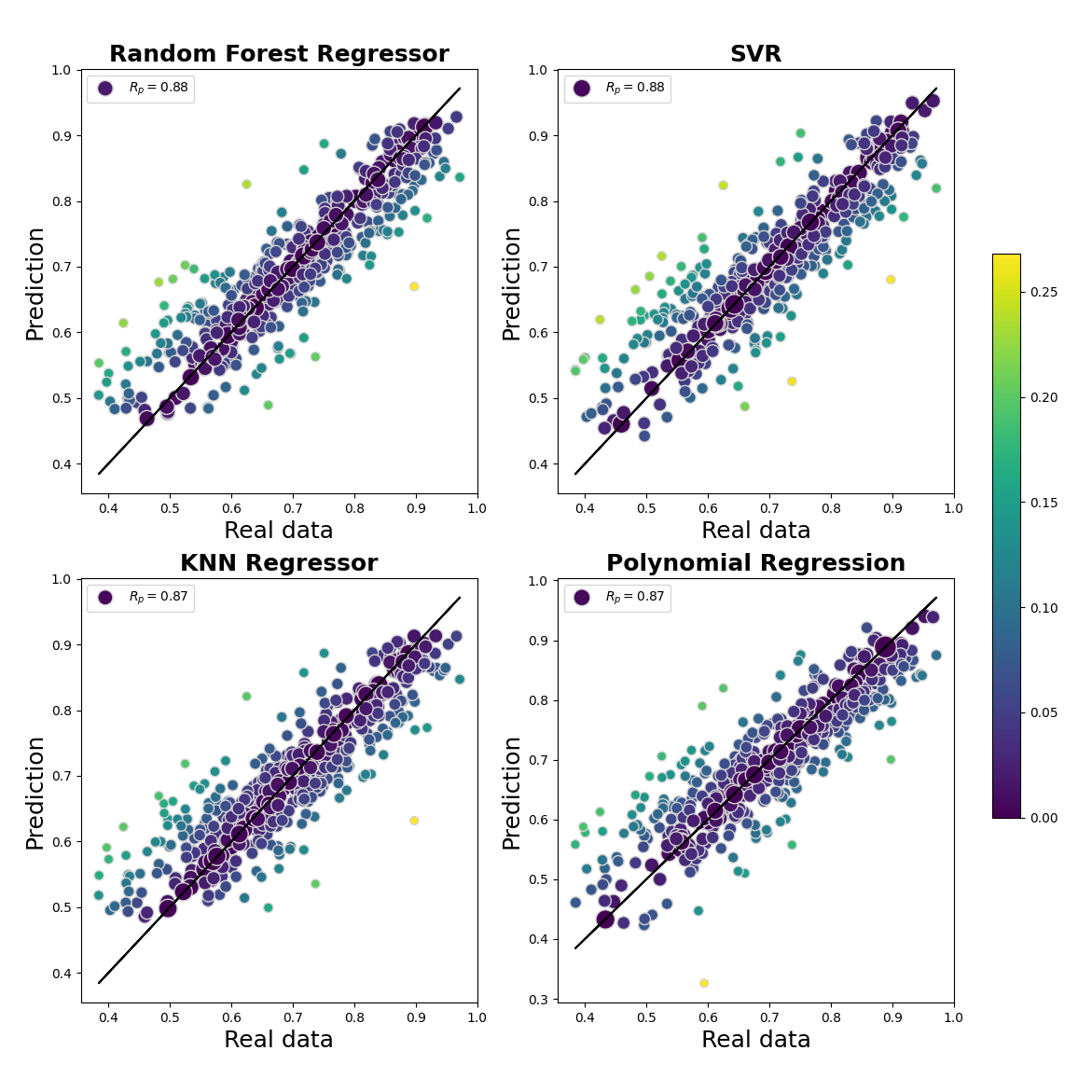}

    \caption{Scatter plots of predicted versus real values for the Random Forest, SVR, KNN, and Polynomial regression models. The black line represents the one-to-one line. The color of data points is an indicator of the absolute difference between real and predicted data ($\rm |y_{pre}-y_{real}|$).} %The size of the data points represents thirty times the logarithm of the inverse of this difference ($\rm 30\times log(1/|y_{pre}-y_{real}|)$).}
    \label{fig:regression_scatter}
\end{figure*}

The scatter plots of the predicted values by these four regression models versus the real data are indicated in Fig.~\ref{fig:regression_scatter}. These plots show that the predicted and real values are in a near one-to-one relation, and the results of the prediction by our regressors are in agreement with the real data. The $R_p$ in Fig.~\ref{fig:regression_scatter}, is the Pearson correlation coefficient, which shows the strength of the linear relationship between two variables, ranging from $-1$ to $+1$. $R_p=+1$ indicates the perfect positive linear relationship. We obtained $R_p$ values between 0.87 and 0.89, indicating a strong correlation between the predicted and actual data.

\section{Summary and Conclusions}\label{summary}
In this paper, we employ machine learning approaches to distinguish pure AGNs from star-forming galaxies, providing a faster alternative to traditional methods. The galaxy sample used in this work is the Galaxy Zoo 1 dataset from SDSS in redshifts of 0.02 to 0.05. This dataset contains 47,675 galaxies, of which 17,002 are pure star-forming galaxies, and 2,254 are active galaxies. To train our machine learning models, we used some significant properties of host galaxies, including redshift, velocity dispersion, stellar mass, [OIII] luminosity, and colors derived from the difference in magnitude. We applied four classification algorithms, using the labels from the BPT (Baldwin, Phillips, and Terlevich) diagnostic diagram as the target for our supervised learning. For this aim, we used the Decision Tree, Random Forest, Support Vector Machine classifier (SVC), and K-Nearest Neighbors (KNN). Based on the obtained value of evaluation metrics (accuracy, precision, recall, and F1-score) and the confusion matrix results, the SVC, followed by Random Forest, achieved the best performance in classifying our galaxies. The evaluation metrics for the SVC and Random Forest are all above 90\%. The ROC result further supports this conclusion. The learning curves indicate that the training processes for SVC, Random Forest, and Decision Tree models do not exhibit overfitting. In contrast, all the mentioned methods to evaluate the models consistently indicate that the KNN classifier performed the worst.

Beyond classification, we aimed to estimate the mass of supermassive black holes ($\rm M_{BH}$) at the nuclei of active galaxies using regression algorithms, as this is a key parameter in studying galaxy evolution. Based on stellar velocity dispersion ($\sigma$) and $\rm M_{BH}-\sigma$ relation, we obtained the logarithm of $\rm M_{BH}/M{\odot}$. We employed Random Forest Regressor, Support Vector Regressor (SVR), KNN Regressor, and Polynomial Regression to predict this property through machine learning algorithms. All models yielded acceptable results with $\rm R^2$ around 0.76. %Among these models, Random Forest Regressor indicates the best result with $\rm R^2 \simeq 0.8$. 
We plotted the predicted values against the true values for four regressors. All scatter plots exhibit distributions close to the one-to-one line, with Pearson correlation coefficients ($\rm R_p$) around 0.9. 

Finally, while the BPT diagram is physically well-motivated and widely used, it requires reliable measurements of specific emission lines and can be limited in cases of low signal-to-noise spectra or missing lines. In contrast, machine learning methods can incorporate a broader set of features simultaneously, model non-linear decision boundaries, and remain applicable even when some spectral information is incomplete. We emphasize that our goal is not to replace the BPT framework, but rather to demonstrate that machine learning techniques provide a flexible and scalable alternative that can complement traditional diagnostics, particularly for large surveys or low-quality data.

\section*{Acknowledgements}
The authors would like to thank the referee for their constructive report and insightful comments, which significantly improved the clarity and quality of this manuscript.

F. M acknowledges support and resources provided by the School of Astronomy at the Institute for Research in Fundamental Sciences-IPM.

%The Acknowledgements section is not numbered. Here you can thank helpful
%colleagues, acknowledge funding agencies, telescopes and facilities used etc.

\section*{Data Availability}
The data underlying this article are available from the Galaxy Zoo public archive at https://data.galaxyzoo.org/\#section-4.



\bibliographystyle{mnras}
\bibliography{reference} % if your bibtex file is called example.bib

%\begin{thebibliography}{99}
%\end{thebibliography}

%%%%%%%%%%%%%%%%%%%%%%%%%%%%%%%%%%%%%%%%%%%%%%%%%%

%%%%%%%%%%%%%%%%% APPENDICES %%%%%%%%%%%%%%%%%%%%%

%\appendix


%If you want to present additional material which would interrupt the flow of the main paper, it can be placed in an Appendix which appears after the list of references.

%%%%%%%%%%%%%%%%%%%%%%%%%%%%%%%%%%%%%%%%%%%%%%%%%%

% Don't change these lines
\bsp	% typesetting comment
\label{lastpage}
\end{document}